\documentclass[pre,aps,twocolumn,10pt,eqsecnum,showpacs]{revtex4-1}

\usepackage[dvips]{graphicx}
\usepackage{textcomp}
\usepackage{amssymb}

\usepackage{mathptmx} 
\usepackage{graphicx,color}

\begin{document}

\title{Delta expansion at low temperatures}
\author{Hirofumi Yamada}\email{yamada.hirofumi@it-chiba.ac.jp}
\affiliation{
Division of mathematics and science, Chiba Institute of Technology, 
\\Shibazono 2-1-1, Narashino, Chiba 275-0023, Japan}

\date{\today}

\begin{abstract}
In the low-temperature phase of a square Ising model, we describe the inverse temperature $\beta$ as a function of the squared mass $M$ and study the critical behavior of $\beta(M)$ via the large $M$ expansion. Using the $\delta$ expansion by which the large mass expansion is transformed into a series exhibiting expected scaling behavior, we estimate the critical inverse temperature $\beta_{c}$ with the help of a linear differential equation to be satisfied by the ansatz of $\beta(M)$. To improve the estimation accuracy, we independently estimate the leading correction exponent $\nu$ from $\beta^{(2)}/\beta^{(1)}$ and then use it to estimate $\beta_{c}$, which significantly improves the accuracy.
\end{abstract}


\maketitle
\section{Introduction}
The $\delta$ expansion introduced in \cite{yam} has been applied to various models and developed to the level that it is now effective in the estimation of critical exponents in the cubic Ising model \cite{yam2}. Due to space-time discretization, the field theoretic and condensed matter models share a similar form of action and allow us to use similar method of calculations. The Boltzmann distribution usually takes the form of $\exp(-\beta H)$, where $H$ means the energy function in a Euclidean lattice and $\beta$ stands for inverse bare coupling for field theory models or inverse temperature in condensed matter models.

In models featuring separated phases, the playing field of the $\delta$ expansion has so far been limited to regions of small $\beta<\beta_{c}$, where $\beta_{c}$ denotes the critical inverse temperature (or critical inverse coupling). In this study, we apply the $\delta$ expansion method to a system in the low-temperature phase for the first time. As an explicit example, we take up a 2D square Ising model. The square Ising model has a long history in the works of Ising, Onsager, Yang, and Lee, as well as many successors \cite{mc}. Making use of exact results, we explore the $\delta$ expansion applied to a new arena of the low-temperature phase. In this paper, we confine ourselves to the very basic points of $\delta$ expansion, such as the estimation of $\nu$ and $\beta_{c}$ from low-temperature expansion. Specifically, we place emphasis on how to independently estimate $\nu$ to improve the $\beta_{c}$ estimate.

This paper is organized as follows.: In section II, we introduce the estimation task by explaining low-temperature expansion and the basic strategy of our approach. In section III, we turn to the explicit estimation of $\beta_{c}$ and the exponent $\nu$, where we make an effort to improve the accuracy of the $\beta_{c}$ estimation. We conclude in section IV with a brief summary.

\section{Low-temperature expansion and basic strategy for estimation}
The square Ising model is controlled by the Boltzmann distribution: 
\begin{equation}
\exp(-\beta\sum_{<i,j>}s_{i}s_{j}),
\end{equation}
where the spin variable $s_{n}\in\{+1, -1\}$ is on the site $n$ that composes the square lattice system. Here, $<i,j>$ means that sites $i$ and $j$ are the nearest neighbor pair and the summation should be taken over all nearest neighbor pairs. 

As in the high-temperature phase, the squared mass is extracted from the two-point function at a large enough separation. At low temperature, there is non-vanishing magnetization per site, ${\cal M}=<s_{0}>$, and the two point function fluctuates around ${\cal M}^2$. At present, the large separation limit of the fluctuation is explicitly known for cases in which the two sites are on diagonal or parallel lines to axes \cite{mc,mont}. Here, we employ the latter case, the same as was considered in \cite{yam3}. The correlation length $\xi$ of the fluctuation is then known as \cite{mont,mc}
\begin{equation}
\xi^{-1}=\log(\tanh\beta)+2\beta.
\label{xi}
\end{equation}
At low enough temperatures, $\xi^{-1}=2\beta-2 e^{-2\beta}-2e^{-6\beta}/3+\cdots$. Since for the application of $\delta$ expansion the corresponding mass squared $M$ proves to be more convenient than $\xi$, we use $M$ defined by \cite{tarko}
\begin{equation}
M=2(\cosh \xi^{-1}-1).
\label{mass}
\end{equation}
From (\ref{xi}) and (\ref{mass}), the squared mass $M$ can be expanded as $M=e^{2\beta}-4+3 e^{-2\beta}+4 e^{-6\beta}+4 e^{-10\beta}+\cdots$. Then, inversion gives
\begin{equation}
\beta=\frac{1}{2}\log M+\frac{2}{M}-\frac{11}{2M^2}+\frac{68}{3M^3}-\frac{451}{4M^4}+\cdots.
\label{largemass}
\end{equation}
This is large mass expansion at low temperature.  It differs compared to the high temperature case in that a logarithmic term of the mass squared exists as the leading contribution. 

The behavior of $\beta(M)$ near the critical point $\beta_{c}=\log(1+\sqrt{2})/2$, given from (\ref{xi}) by solving $\xi^{-1}=0$, is easily derived from (\ref{xi}) and (\ref{mass}). The result of expansion provides
\begin{equation}
\beta=\beta_{c}+\frac{M^{1/2}}{4}(1-\frac{M}{24}+\frac{3M^2}{640}-\cdots)+R,
\label{beta_scaling}
\end{equation}
where $R$ denotes the analytic background given by
\begin{equation}
R=\frac{M}{16\sqrt{2}}(1-\frac{3M}{32}+\frac{19M^2}{1536}-\cdots).
\end{equation}
From the definition of critical exponent $\nu$, $\xi\sim (\beta-\beta_{c})^{\nu}$ and $\beta-\beta_{c}\sim M^{1/2\nu}$. Then, we find from (\ref{beta_scaling}) that $\nu=1$.

The motivation behind our recent series of works is to pave the way for a new quantitative computational method of critical behaviors from simply accessible series expansions, such as high- and low-temperature expansions. The $\delta$-expansion \cite{yam,yam3} provides us with a new way of handling the series and enables the estimation of critical quantities. Suppose a given truncated expansion of $f(M)$ to the order $N$, $f(M)=\sum_{n=0}^{N}a_{n}(1/M)^n$. The minimal result of $\delta$-expansion needed in this study is that it induces transformation summarized as
\begin{equation}
D_{N}[M^{-\lambda}]=C_{N,\lambda}t^{\lambda},\label{delta}
\end{equation}
where
\begin{equation}
C_{N,\lambda}=\frac{\Gamma(N+1)}{\Gamma(\lambda+1)\Gamma(N-\lambda+1)}.
\end{equation}
Here, $D_{N}$ means the transformation that is $N$ dependent. Note that $D_{N}[1]=1$ and $D_{N}[M^{\ell}]=0$ for positive integer $\ell$. Then, we have
\begin{equation}
D_{N}[f]=:\bar f(t)=\sum_{n=0}^{N}C_{N,n}a_{n}t^n.
\end{equation}
The coefficients are dependent on the truncation order $N$. In the examples investigated so far, it is numerically verified that $\bar f(t)$ recovers in its effective region the small $M$ behavior of $f(M)$ \cite{comment1}. 

In the low-temperature phase, $\beta(M)$ at large $M$ involves $\log M$ as manifested in (\ref{largemass}). The transformation rule of the logarithmic function can also be drawn from (\ref{delta}). For instance, putting $\lambda=\epsilon$ and expanding in $\epsilon$ and the comparison of coefficients of $\epsilon$ in the result, we find
\begin{equation}
D_{N}[\log M]=-\log t-\sum_{n=1}^{N}\frac{1}{n}.
\end{equation}

In what follows, we use notation $f_{>}$ for the series expansion of $f$ at large $M$. Expansion at small $M$ is denoted as $f_{<}$. The transformed series follows the same notation. For example, we obtain
\begin{eqnarray}
D_{N}[\beta>]&=:&\bar\beta_{>}(t)=\frac{1}{2}(-\log t-\sum_{k=1}^{N}\frac{1}{k})\nonumber\\
& +&C_{N,1}2t-C_{N,2}\frac{11}{2}t^2+C_{N,3}\frac{68}{3}t^3-\cdots,
\end{eqnarray}
where the last term is $const\times t^N$. We now see the behaviors of $\bar\beta_{>}$ and its derivatives with respect to $\log t$ in Fig. 1. 
\begin{figure}
\centering
\includegraphics[scale=0.9]{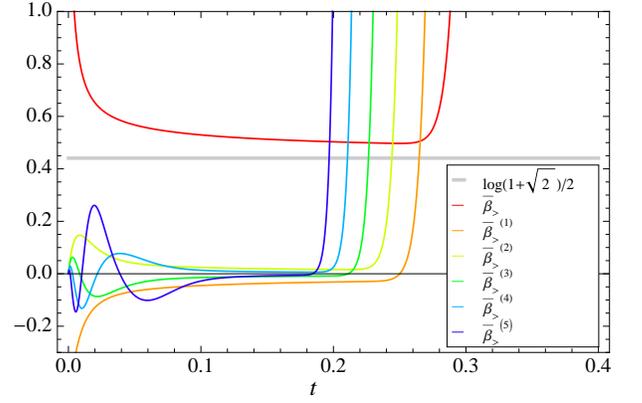}
\caption{The plots of $\bar \beta_{>}$ at $N=25$ and its derivatives with respect to $d/d\log t$ to the 5th order. The gray line indicates $\beta_{c}=\log(1+\sqrt{2})/2=0.4406867935\cdots$.}
\end{figure}
From the plots, we find that $\bar\beta_{>}$ gradually approaches $\beta_{c}$ but the speed of convergence is too slow. This is because $M^{1/2\nu}$ exists in the second place of the expansion (\ref{beta_scaling}). If we could effectively subtract the correction terms, an accurate estimation of $\beta_{c}$ would be possible. This strategy is conveniently carried out by setting up the differential equation approximately satisfied by $\beta_{<}$ \cite{yam3,yam2}. 

Suppose we have no information on the exponents and let $\beta_{<}=\beta_{c}+\sum_{n=1}A_{n}M^{\lambda_{n}}$. Then, the $\delta$-expansion to order $N$ is given by 
\begin{equation}
\bar \beta_{<}=\beta_{c}+\sum_{n=1}C_{N, -p_{n}}A_{n}t^{-p_{n}}.
\label{ansatz1}
\end{equation}
Here, the set of exponents $\{p_{n}\}$ is the subset of $\{\lambda_{n}\}$ obtained from removing the integer ones (Note that $D_{N}[M^{\ell}]=0$, $(\ell=1,2,3,\cdots)$. The analytic part $R$ thus becomes negligible.). Truncating the summation at order $K$ in (\ref{ansatz1}), the resulting series $\bar \beta_{<}=\beta_{c}+\sum_{n=1}^{K}C_{N, p_{n}}A_{n}t^{-p_{n}}$ is  considered ansatz to the $K$th order. It satisfies
\begin{equation}
\prod_{k=1}^K[1+p_{k}^{-1}\frac{d}{d\log t}]\bar \beta_{<}=\beta_{c}.
\end{equation}
In this linear differential equation (LDE), the highest derivative order is $K$. If the transformed function $\bar\beta_{>}^{(K)}$ shows expected scaling, it is allowed to substitute $\bar\beta_{>}^{(n)}$ ($n=0,1,2,\cdots, K$) into $\bar\beta_{<}^{(n)}$ included in the above LDE.

Now, in reality, LDE is only valid locally at a certain $t$ due to the truncation of expansion. Then, at $te^{\epsilon}$ in the neighborhood of $t$, we have the following expansion:
\begin{eqnarray}
\prod_{k=1}^K[1+p_{k}^{-1}\frac{d}{d\log t}]\bar \beta(t e^{\epsilon})
&=&\prod_{k=1}^K[1+p_{k}^{-1}\frac{d}{d\log t}]\beta(t)\nonumber\\
& &+\prod_{k=1}^K[1+p_{k}^{-1}\frac{d}{d\log t}]\beta^{(1)}(t)\epsilon
\nonumber\\
& &+O(\epsilon^2).
\end{eqnarray}
A good adjustment of the values of unknown exponents $p_{1}, p_{2},\cdots$ should make LDE approximately valid over a wide region of a plateau indicating $\beta_{c}$. Thus, we employ the extended version of the principle of minimum sensitivity \cite{steve} (PMS) to fix the exponents and $t$ at which $\beta_{c}$ is estimated. The extended PMS reads
\begin{equation}
\prod_{k=1}^K[1+p_{k}^{-1}\frac{d}{d\log t}]\bar \beta_{<}^{(n)}=0
\end{equation}
for various sets of $n$. In the above LDE, unknown variables are $(t; p_{1},p_{2},\cdots, p_{K})$. Hence, we need $K+1$ equations to estimate all values. Then, the last LDE includes a $2K+1$th order derivative. At low orders, fewer derivatives exhibit scaling, and at large orders, derivatives to several orders are available. When the solution of the set $(t^{*}; p_{1}^{*},\cdots,p_{K}^{*})$ is obtained in the scaling region, we estimate $\beta_{c}$ by
\begin{equation}
\prod_{k=1}^K[1+p_{k}^{-1}\frac{d}{d\log t}]\bar \beta_{<}\Big|_{(t^{*}; p_{1}^{*},\cdots,p_{K}^{*})}=\beta_{c}.
\label{betaestimate}
\end{equation}
This is the basic strategy utilized in our approach. A similar approach is taken for the estimation of $\nu$ and the improved estimation of $\beta_{c}$.

\section{Estimation}
\subsection{Naive estimation of $\beta_{c}$}
In this subsection, we present our estimation study in the manner detailed in the previous section. In the protocol, we treat introduced exponents as adjustable ones to satisfy the extended PMS condition, which was adopted in \cite{yam4}. Here, we call this protocol the naive protocol. 

The number of exponents in ansatz (\ref{ansatz1}) agrees with the order $K$ of LDE ($k$LDE) to be imposed. At $1$LDE, we use ansatz $\beta_{<}=\beta_{c}+A_{1}M^{p_{1}}$ and impose $[1+p_{1}^{-1}(d/d\log t)]\bar\beta_{<}=\beta_{c}$. Then, we first consider
\begin{equation}
[1+p_{1}^{-1}(d/d\log t)]\bar\beta_{<}^{(k)}=0,\quad k=1,2.
\end{equation} 
This set of LDEs includes derivatives to the 3rd order. By substituting $\bar\beta_{>}(t)$ into $\bar\beta_{<}(t)$, we obtain the solution of the set $(t^{*}; p_{1}^{*})$. For details on the estimation process, see \cite{yam3,yam4}. Then, by substituting set $(t^{*}; p_{1}^{*})$ into the left-hand-side of $[1+p_{1}^{-1}(d/d\log t)]\bar\beta_{<}=\beta_{c}$, we obtain the estimation of $\beta_{c}$, as suggested in (\ref{betaestimate}). We also performed for $2$- and $3$-parameter ansatz the same estimation protocol to the $50$th order. The result is given in Fig. 2 and summarized in Table I.
\begin{figure}
\centering
\includegraphics[scale=0.9]{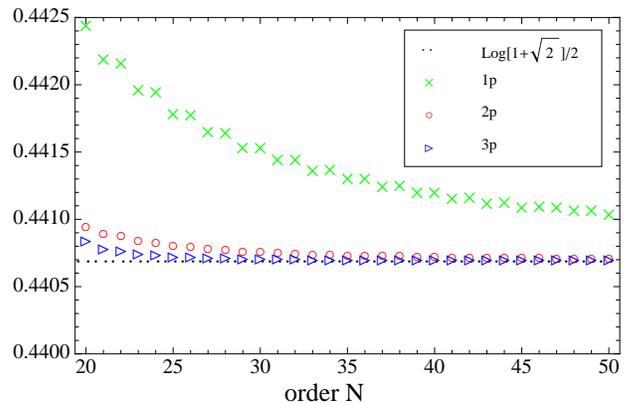}
\caption{Plots of naive estimation results with $K$-parameter ansatz where $K=1,2,3$.}
\end{figure}
\begin{table*}[b]
\caption{Naive $K$- parameter estimation results of $\beta_{c}=0.4406867935\cdots$ via $\bar \beta$ under the naive protocol.}
\begin{center}
\begin{tabular}{lccccccc}
\hline\noalign{\smallskip}
& $20$ & $25$ & $30$ & $35$ & $40$ & $45$ & $50$\\
\noalign{\smallskip}\hline\noalign{\smallskip}
$K=1$  & 0.44243483 & 0.44178017 & 0.44152678  &  0.44129454 & 0.44119466  & 0.44108312 & 0.44103340 \\
$K=2$  &  0.44093637 &  0.44079480 & 0.44075068 & 0.44072337 & 0.44071252 & 0.44070368 & 0.44069980  \\
$K=3$  & 0.44083163  &  0.44071513  & 0.44069812 & 0.44069176 & 0.44068964 & 0.44068834 & 0.44068783 \\
\noalign{\smallskip}\hline
\end{tabular}
\end{center}
\end{table*}

The scaling behavior of the left-hand side of LDE sets in at orders depending on the number of exponent parameters of ansatz (in this context, see \cite{yam3,yam4}). At $K=1$, it is no sharp transition for odd orders. For even orders, rough scaling behavior begins to appear from the $6$th order. At $K=2$, the onset orders are $11$th for odd orders and $14$th for even orders. At $K=3$, the orders are $19$th for odd $N$ and $22$nd for even $N$. 

As expected, increasing the number of terms in the ansatz improves the accuracy. However, the reliability of estimates for each $K$ actually depends on the order $N$ of large mass expansion, as suggested just before by the concerned function's behaviors. Although the convergence to the exact value of $\beta_{c}$ is strongly indicated, at least with $2$- and $3$-parameters, we would still like to improve the accuracy. As in the high-temperature case demonstrated in \cite{yam3}, we turn to the estimation of $\nu$ and then revisit the estimation of $\beta_{c}$ under the bias of the estimated $\nu$.

\subsection{Estimation of $\nu$}
As long as the order $N$ of a large mass expansion is big enough, the estimation of $\nu=(2p_{1})^{-1}$ is effective in the naive estimation discussed above. However, a precise value cannot be obtained at moderate orders. It seems that obtaining a precise value of $p_{1}$ would help to estimate $\beta_{c}$ since the leading correction $t^{-p_{1}}$ is effectively subtracted in LDE. To solve this problem, we consider the function $\beta^{(2)}/\beta^{(1)}=:f_{\beta}$, where $\beta^{(\ell)}(M)=(d/d\log M)^{\ell}\beta(M)$. From (\ref{beta_scaling}), we find the critical behavior
\begin{equation}
f_{\beta<}(M)=\frac{1}{2\nu}+\frac{M^{1/2}}{4\sqrt{2}}-\frac{3M}{16}+\cdots.
\end{equation}
The exponent $p_{1}$ appears as the leading term of the constant and its estimation may become accurate. Writing ansatz as $f_{\beta<}(M)=p_{1}+\sum_{k=1}B_{k} M^{q_{k}}$, we obtain its $\delta$ expansion, giving
\begin{equation}
\bar f_{\beta<}(t)=p_{1}+\sum_{k=1}C_{N, q_{k}}B_{k} t^{-q_{k}}.
\end{equation}
The truncated series at $t^{-q_{K}}$ satisfies the following LDE:
\begin{equation}
\prod_{k=1}^K[1+q_{k}^{-1}\frac{d}{d\log t}]\bar f_{\beta<}=p_{1}.
\label{lde1}
\end{equation}
As in the previous section, the above LDE should be valid at certain $t$ and the estimation of $p_{1}$ requires the optimal set $(t^*; q_{1}^*,\cdots, q_{K}^*)$. The set is determined by the extended PMS.
\begin{figure}
\centering
\includegraphics[scale=0.9]{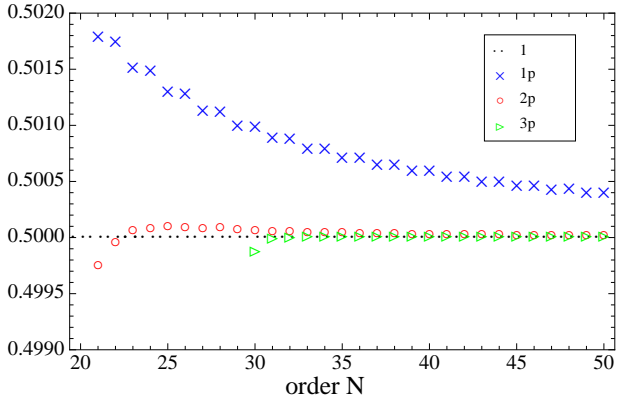}
\caption{Plots of $p_{1}=1/(2\nu)=0.5$ estimate. $1$p ($2$p, $3$p) denotes the result in the $1$ ($2$, $3$)-parameter ansatz.}
\end{figure}

Now, we attempt to estimate $p_{1}$ from the large mass expansion obtained from (\ref{largemass})
\begin{equation}
f_{\beta>}(x)=\frac{4}{M}-\frac{28}{M^2}+\frac{208}{M^3}-\frac{1616}{M^4}+\cdots.
\end{equation}
We substitute $\bar f_{\beta>}$ into $\bar f_{\beta<}$ in the derivatives of (\ref{lde1}), $\prod_{k=1}^K[1+q_{k}^{-1}(d/d\log t)]\bar f_{\beta<}^{(\ell)}=0$ $(\ell=1\sim K+1)$. The solution determines the optimal set $(t^*; q_{1}^*,\cdots, q_{K}^*)$ and the substitution into (\ref{lde1}) provides the $p_{1}$ estimate. The result is shown in Fig. 3 and summarized in Table II.
\begin{table*}
\caption{$K_{\nu}$- and $K$- parameter estimation results of $p_{1}=0.5$ via $\bar f_{\beta}$ and $\bar \beta$, respectively. At $K_{\nu}=1$, a proper estimate appears from the $17$th for odd orders and the $20$th for even orders. At $K_{\nu}=2$, a proper estimate appears from the $27$th for odd orders and the $30$th for even orders. At $K_{\nu}=3$, a proper estimate appears from the $37$th for odd orders and the $42$nd for odd orders. }
\begin{center}
\begin{tabular}{lccccccc}
\hline\noalign{\smallskip}
& $20$ & $25$ & $30$ & $35$ & $40$ & $45$ & $50$\\
\noalign{\smallskip}\hline\noalign{\smallskip}
$K_{\nu}=1$  & 0.50208109  & 0.50129729  &  0.50098433  &  0.50071069 & 0.50058813  & 0.50045795 & 0.50039751 \\
$K=1$  & 0.51821317 & 0.51317267  &  0.51100350  &  0.50882559 & 0.50781555  & 0.50660710 & 0.50603627 \\
$K_{\nu}=2$  &  0.49843886 &  0.50008775 & 0.50006051 & 0.50003490 & 0.50002454 & 0.50001621 & 0.50001244  \\
$K=2$  &  0.50380104 &  0.50199096 & 0.50132581 & 0.50085852 & 0.50065241 & 0.50046897 & 0.50038216  \\
$K_{\nu}=3$  &    &    &  0.50058851 & 0.50000241 & 0.50000213 & 0.50000098 & 0.50000066 \\
$K=3$  & 0.50243597  &  0.50064853  & 0.50030467 & 0.50015361 & 0.50009665 & 0.50005811 & 0.50004131 \\
\noalign{\smallskip}\hline
\end{tabular}
\end{center}
\end{table*}

The quality of the series of $\bar f_{\beta}$ is lower than that of $\bar\beta$. In the $1$-parameter ansatz, the orders at which the solution $(t^*; q_{1}^*)$ is found at the rough scaling region (we call such solutions gproperh) are $17$th for odd order and $20$th for even order. In the $2$-parameter ansatz, the behavior characteristic to $2$-parameter ansatz appears at the $19$th (for odd) and the $20$th (for even) orders and proper solution sets in from the $27$th (for odd) and the $30$th (for even) in the $3$-parameter ansatz.

This quality difference affects the estimation accuracy at low orders. For the same given orders, the estimate from $\bar f_{\beta}$ is slightly more accurate than the estimate from $\bar\beta$. As the order grows, the estimate from $\bar f_{\beta}$ improves more and more. As for the relation between the accuracy and the number of parameters, a positive correlation is clearly confirmed. The main finding of this experiment is that the estimation of critical exponent $\nu$ is better in $f_{\beta}$.

\subsection{Improved estimation of $\beta_{c}$}
In this subsection, we experimentally investigate the best way to estimate $\beta_{c}$. In addition to the naive way presented in the subsection A, we here examine two ways: (i) using $p_{1}$ estimated from $\bar f_{\beta}$, with the rest of the exponents adjusted under the extended PMS as usual, and (ii) using exactly known values of exponents $p_{i}$ for all exponents included in the ansatz. For example, at $K=2$, we simply substitute $p_{1}=1/2$, $p_{2}=3/2$, and at $K=3$, we substitute in addition $p_{3}=5/2$. This prescription supplies us  the standard reference of estimation accuracy.

\begin{figure}
\centering
\includegraphics[scale=0.9]{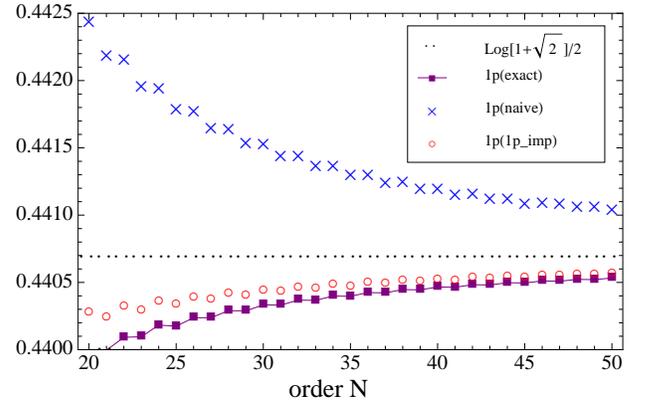}
\caption{Plots of $\beta_{c}$ estimates. $1$p (exact) means the result under the substitution $p_{1}=1/2$ (prescription (ii)). $1$p (naive) means the result under the naive prescription. $1$p ($1$p-imp) means the result under the substitution $p_{1}=p_{1}^{*}$, where $p_{1}^*$ is estimated in $1$-parameter ansatz of $\bar f_{\beta<}$ (prescription (i)).}
\end{figure}
\begin{figure}
\centering
\includegraphics[scale=0.9]{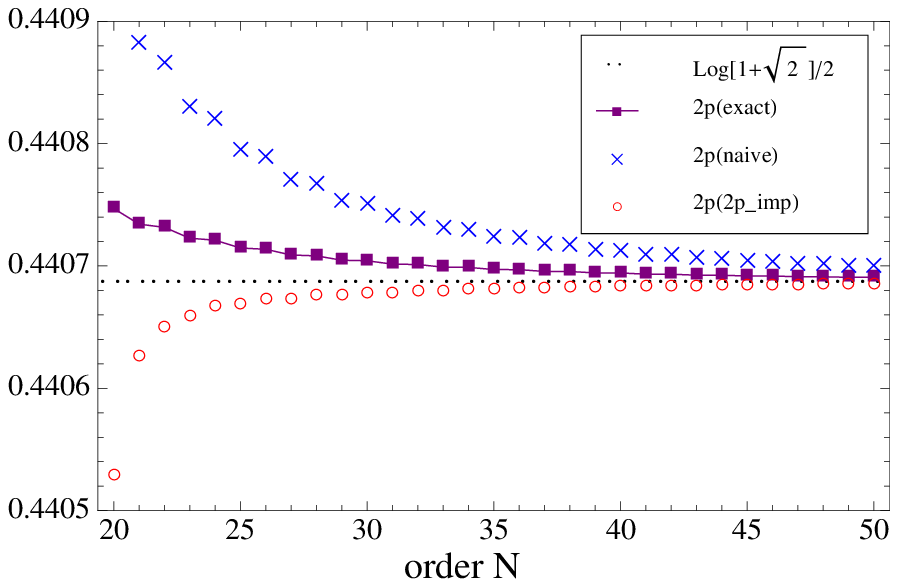}
\caption{Plots of $\beta_{c}$ estimates. $2$p (exact) means the result under substitution $p_{1}=1/2$ (prescription (ii)). $2$p (naive) means the result under the naive prescription. $2$p ($2$p-imp) means the result under substitution $p_{1}=p_{1}^{*}$, where $p_{1}^*$ is estimated in the $2$-parameter ansatz of $\bar f_{\beta<}$ (prescription (i)).}
\end{figure}
\begin{figure}
\centering
\includegraphics[scale=0.9]{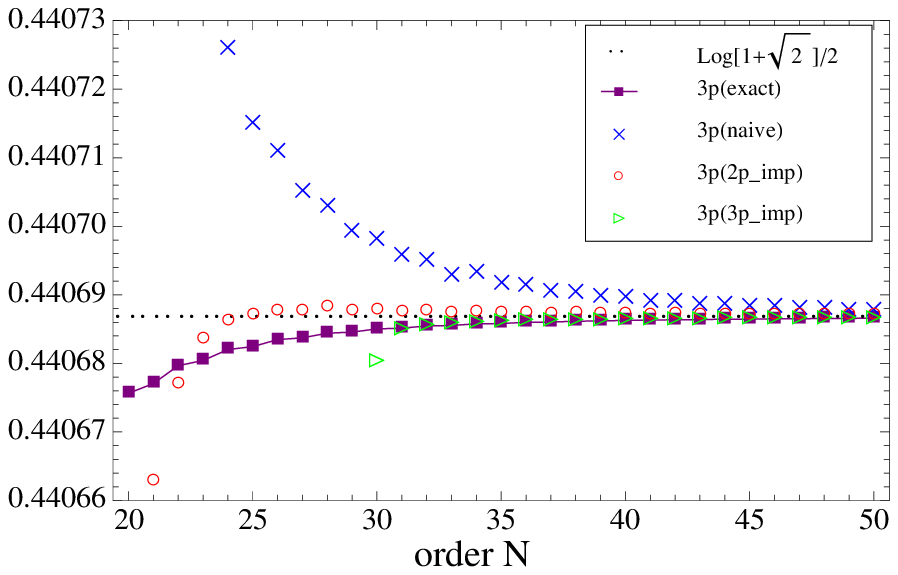}
\caption{Plots of $\beta_{c}$ estimates. $3$p (exact) means the result under substitution $p_{1}=1/2$ (prescription (ii)). $3$p (naive) means the result under the naive prescription. $3$p ($k$p-imp, $k=1,2$) means the result under substitution $p_{1}=p_{1}^{*}$, where $p_{1}^*$ is estimated in the $k$-parameter ansatz of $\bar f_{\beta<}$ (prescription (i)).}
\end{figure}
As shown in Fig. 4, in the $1$-parameter cases, both prescriptions (i) and (ii) provide better estimates than the naive one, with (i) yielding the best estimates among the three. In the $2$-parameter cases, we first note that the used $p_{1}$ value is the one from the $2$-parameter ansatz of $\bar f_{<}$ (see Fig. 5). The result is that the whole trend is the same as that of the $1$-parameter ansatz. Still, we note that prescription (i) becomes the best one from the $22$nd order. In the $3$-parameter cases, we examined two versions of prescription (i): one, using $p_{1}$ from the $2$-parameter ansatz of $\bar f_{<}$, and two, using $p_{1}$ from the $3$-parameter ansatz of $\bar f_{<}$. In the latter case, stable estimation sets in from the $30$th order (see Fig. 6). In $2$- and $3$- parameter ansatz, prescriptions (i) and (ii) resulted in competing estimates, and numerical Table III is necessary for discussion. 

From Table III, we find in the $2$-parameter case that from moderate orders, or as the order grows higher, the best estimation is realized in prescription (i), where $p_{1}^*$ at $K_{\nu}=2$ is substituted and $p_{2}$ is adjusted to satisfy extended PMS. A similar trend is observed in the $3$-parameter case, though the $50$th order is exceptional. Though the exact reason is not known, we can conclude that the best estimation does not come from the substitution of exact exponents in the ansatz; rather, the use of the leading exponent value obtained from $\bar f_{\beta}$ with extended PMS provided the best results.

\begin{table*}
\caption{Relative error ($\beta_{c,estimated}/\beta_{c}-1$) in $2$- parameter estimation results of $\beta_{c}=0.4406867935\cdots$ via three prescriptions.}
\begin{center}
\begin{tabular}{lrrrrrrr}
\hline\noalign{\smallskip}
& $20$ & $25$ & $30$ & $35$ & $40$ & $45$ & $50$\\
\noalign{\smallskip}\hline\noalign{\smallskip}
$Naive$  & 0.00056634 & 0.00024509  & 0.00014498  & 0.00008301 & 0.00005839  & 0.00003832 & 0.00002952 \\
$Exact$  & 0.00013657 & 0.00006330  & 0.00003965  & 0.00002339 & 0.00001698  & 0.00001135 & 0.00000893  \\
$p_{1}(K_{\nu}=2)$  & -0.00035963  &  -0.00004278  & -0.00002146 & -0.00001539 & -0.00000959 & -0.00000751 & -0.00000517 \\
\noalign{\smallskip}\hline
\end{tabular}
\end{center}
\end{table*}

\begin{table*}
\caption{Relative error ($\beta_{c,estimated}/\beta_{c}-1$) in $3$- parameter estimation results of $\beta_{c}=0.4406867935\cdots$ via three prescriptions.}
\begin{center}
\begin{tabular}{lrrrrrrr}
\hline\noalign{\smallskip}
& $20$ & $25$ & $30$ & $35$ & $40$ & $45$ & $50$\\
\noalign{\smallskip}\hline\noalign{\smallskip}
$Naive$  & 0.000328649 & 0.000064305  & 0.000025697  & 0.000011261 & 0.000006458  & 0.000003519 & 0.000002342 \\
$Exact$  & 0.000025312 & 0.000009856  & 0.000003990  & 0.000002217 & 0.000001156  & 0.000000759 & 0.000000455  \\
$p_{1}(K_{\nu}=2)$  & -0.000216960  &  0.000000710  & 0.000002244 & 0.000001161 & 0.000000944 & 0.000000567 & 0.000000464 \\
$p_{1}(K_{\nu}=3)$ & &  & -0.000014723  & -0.000001518  & -0.000000711 & -0.000000483 & -0.000000282  \\
\noalign{\smallskip}\hline
\end{tabular}
\end{center}
\end{table*}

\section{Summary}
We performed the basic task of estimating critical quantities $\nu$ and $\beta_{c}$ and demonstrated that the large mass expansion with $\delta$ expansion also works at the low-temperature phase. We also found that prescription (i), the substitution of leading correction exponent $p_{1}=1/(2\nu)$ estimated with $\bar f_{\beta}$ under extended PMS, provided better $\beta_{c}$ value than naive estimation of $\beta_{c}$ with extended PMS. Moreover, the (i)-prescription delivered better estimates than the prescription in which exact exponents in the ansatz are used. This sheds light on the accurate estimation of $\beta_{c}$ in the cubic model presented in \cite{yam2}.

Encouraged with the results thus far, we intend to attempt computation of the exponent of spontaneous magnetization ${\cal M}$ in the present approach and additional estimation of other critical quantities from the low-temperature phase. By these further studies on the method of large mass expansion with $\delta$-expansion, its full aspects and power will be grasped. Such a thorough examination may stimulate the improvement of the method and help us to apply the method to more complicated and interesting physics models.


\begin{thebibliography}{99}
\bibitem{comment1} As explained in \cite{yam}, the $\delta$ expansion originates in the dilation operation of the scaling region. From this point of view, it is plausible that $\bar f(t)$ is able to capture the scaling behavior.
\end{thebibliography}
\end{document}